# Enhanced Control of Transient Raman Scattering Using Buffered Hydrogen in Hollow-Core PCF

P. Hosseini[1], D. Novoa[1], A. Abdolvand[1], and P. St.J. Russell[1,2]
[1]*Max Planck Institute for the Science of Light, Staudtstraße 2, 91058 Erlangen, Germany*
[2]*Department of Physics, University of Erlangen-Nuremberg, 91058 Erlangen, Germany*

Many reports on stimulated Raman scattering in mixtures of Raman-active and noble gases indicate that the addition of a dispersive buffer gas increases the phase-mismatch to higher-order Stokes/anti-Stokes sidebands, resulting in preferential conversion to the first few Stokes lines, accompanied by a significant reduction in Raman gain due to collisions with buffer gas molecules. Here we report that, provided the dispersion can be precisely controlled, the effective Raman gain in gas-filled hollow-core photonic crystal fiber (PCF) can actually be significantly enhanced when a buffer gas is added. This counterintuitive behavior occurs when the nonlinear coupling between the interacting fields is strong, and can result in a performance similar to that of a pure Raman-active gas, but at much lower total gas pressure, allowing competing effects such as Raman backscattering to be suppressed. We report high modal purity in all the emitted sidebands, along with anti-Stokes conversion efficiencies as high as 5% in the visible and 2% in the ultraviolet. The results point to a new class of gas-based waveguide device in which the pressure-tunable nonlinear optical response is beneficially adjusted by the addition of other gases.

Mixtures of atomic and molecular gases have been useful for improving the efficiency of high-harmonic generation [1], controlling the spatial distribution of multiple filaments [2], and increasing the output power of copper-vapor lasers [3]. It has also been shown that the addition of a noble buffer gas to a Raman-active gas increases the phase-mismatch for the generation of higher-order sidebands in stimulated Raman scattering (SRS), restricting conversion to the first few Stokes lines [4,5]. Most of these studies, carried out in free-space arrangements, were performed in the steady-state regime of SRS, when the Raman gain falls as a result of collisional broadening at higher partial buffer gas pressures. This gain reduction can however be drastically mitigated in gas-filled hollow-core photonic crystal fiber (HC-PCF), which offers long collinear interaction lengths at high pump intensity, permitting operation in the so-called transient SRS regime, when the duration of the pump pulses is comparable to the lifetime of the molecular oscillations $T_2$ [6]. This can lead to the generation of spectral clusters in HC-PCFs filled with gas mixtures [7].

In this Letter we report that the effective nonlinear response of a broadband-guiding HC-PCF, filled with a binary mixture of a molecular and an atomic gas, uniquely allows efficient generation of Raman sidebands, in a pure fundamental guided mode, at wavelengths from the ultraviolet to the near-infrared. In contrast to gas-cells and bulk systems, the Raman gain in the gas itself is insufficient to explain the observed SRS dynamics in gas-filled HC-PCFs. This is because the interplay between the normal dispersion of the gas and the spectrally smooth and anomalous hollow waveguide dispersion strongly affects the net Raman gain when nonlinear coupling between the pump, Stokes and anti-Stokes fields is strong [8,9]. Under suitable conditions, the net Raman gain in a gas-mixture-filled HC-PCF can be higher than in the pure gas, and at much lower total pressures. In a pure gas system, the higher Raman-active gas pressure dramatically modifies the nonlinear dynamics, triggering parasitic effects such as Raman backscattering, which are absent in the gas-mixture-filled HC-PCF. This rather counter-intuitive result requires operation close to a zero dispersion point, which is very difficult if not impossible to arrange in a collinear free-space geometry.

We report a series of experiments on vibrational SRS (frequency shift 125 THz) in which a precisely prepared mixture of $H_2$ and Xe is introduced into a short length of kagomé-style HC-PCF. Pumping the fiber at 532 nm results in conversion efficiencies of 5% to the 435 nm anti-Stokes band and 2% to the 368 nm second anti-Stokes band in the ultraviolet. Remarkably, all the sidebands are emitted in the fundamental $LP_{01}$-like core mode, in sharp contrast to previous studies [10,11].

In the limit of no pump depletion, the steady-state exponential gain factor for Stokes light in the $LP_{01}$-like mode, pumped in the $LP_{01}$-like mode, is $G_{01}^{SS} = \rho_{01} S_{01} g_P I_P L$, where $g_P$ is the vibrational Raman gain in bulk hydrogen ($g_P \propto T_2$) [8], $S_{01}$ is the intermodal overlap integral, $\rho_{01} \leq 1$ is the gain reduction factor [9], $I_P$ the pump intensity and $L$ the fiber length [8]. In the steady-state [12], $g_P$ saturates at pressures above 10 bar, i.e., above the Dicke narrowing pressure [13]. This is however no longer the case if the system operates in the so-called transient regime, which holds when the following conditions are satisfied:

$$g_P I_P L \gg \tau_P / T_2 = \hat{\tau}_P \quad \text{and} \quad g_P I_P L \gg 1/\hat{\tau}_P \qquad (1)$$

where $\tau_P$ is the duration of pump pulse (assumed to have a square profile) and $T_2$ is inversely proportional to the total gas pressure [14]. We have verified that the above conditions are always fulfilled for our experimental parameters [15].

In a molecular-monoatomic gas mixture, the inverse Raman lifetime can be written as:

$$T_2^{-1}[\text{MHz}] = \pi\left(Ap_R^{-1} + Bp_R + Cp_B\right), \quad (2)$$

where $p_R$ and $p_B$ are respectively the partial pressures of the Raman-active and buffer gases, $A$ is the diffusion-limited broadening coefficient, $B$ is the collisional self-broadening coefficient and $C$ is the cross-broadening coefficient. At 298 K the values of $A$ and $B$ are 280 MHz·bar and 57 MHz/bar for the fundamental vibrational transition of $H_2$ [16], and the $C$ values for different $H_2$-rare gas mixtures can be found in [17].

Provided Eq. (1) is satisfied, the power amplification factor of the Stokes signal in bulk gas, again in the limit of no pump depletion, and for a square pump pulse, is [15]:

$$\left|\frac{E_S(L)}{E_S(0)}\right|^2 = \frac{1}{\pi}\frac{\exp\sqrt{8g_P I_P L \hat{\tau}_P}}{\sqrt{8g_P I_P L \hat{\tau}_P}} e^{-2\hat{\tau}_P} \simeq \frac{1}{\pi} e^{\sqrt{8g_P I_P L \hat{\tau}_P} - 2\hat{\tau}_P} \quad (3)$$

where the approximation follows from the second condition in Eq. (1). The transient exponential gain factor in the fiber can then be written:

$$G_{01}^{TR} = \sqrt{8g_P \rho_{01} S_{01} I_P L \hat{\tau}_P} - 2\hat{\tau}_P. \quad (4)$$

Although increasing the buffer gas pressure will increase the second term, for large enough values of $I_P L$ (easily attainable in HC-PCF) the first term will dominate [7]. This means that the system can operate in the transient regime even with pulse durations in the nanosecond range [6].

The dominant effect of a buffer gas is actually to modify the fiber dispersion and thus the gain reduction factor, through $\rho_{01} \propto \vartheta_{01}/g_P I_P$, where $\vartheta_{01} = (\beta_S^{01} + \beta_{AS}^{01}) - 2\beta_P^{01}$ is the dephasing rate and $\beta_J^{01}$ is the propagation constant of the J-th sideband travelling in the $LP_{01}$-like mode [8]. To illustrate this, in Fig. 1 we explore the dependence of $G_{01}^{TR}$ on pump intensity and partial pressure in a kagomé-PCF with 11 μm core radius and 90 nm core-wall thickness, pumped at 532 nm with 1 ns pulses. For these parameters, coherent Raman gain suppression ($\rho_{01} = 0$, i.e., $\vartheta_{01} = 0$) occurs at $p_{H2} \sim 18$ bar [9]. Fig. 1(a) shows that, for fixed $I_P = 15$ W/μm², the pressure at which $\rho_{01} = 0$ can be tuned simply by increasing $p_{Xe}$ at any value of $p_{H2} < 18$ bar. Similarly, when $p_{H2}$ is fixed at 15 bar (Fig. 1(b)), the gain is strongly enhanced with increasing pump intensity, especially in the vicinity of $p_{Xe} \sim 4$ bar.

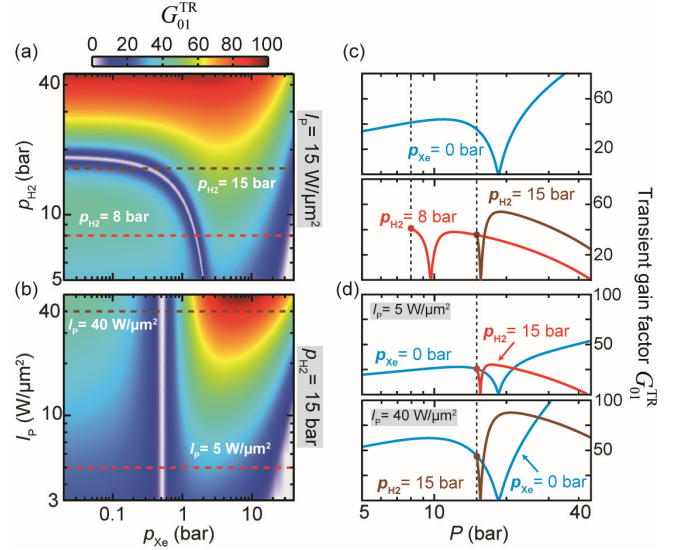

FIG. 1. Transient exponential gain factor $G_{01}^{TR}$ plotted against (a) $p_{H2}$ and $p_{Xe}$ for $I_p = 15$ W/μm² and (b) $I_P$ and $p_{Xe}$ for $p_{H2} = 15$ bar. (c) and (d) are cuts through at the positions of the horizontal dashed lines in (a) and (b), respectively. The blue curves in (c) and (d) show the behavior when $p_{Xe} = 0$.

The lower panel in Fig. 1(c) shows horizontal slices through Fig. 1(a) at $p_{H2} = 8$ and 15 bar (note that they are plotted againsts the total pressure $P = p_{H2} + p_{Xe}$). The upper panel shows the behavior when $p_{Xe} = 0$ (pure hydrogen). It is clear that, whereas $G_{01}^{TR}$ in the gas mixture qualitatively follows the same behavior as pure hydrogen, i.e., the gain drops to zero at $\vartheta_{01} = 0$ and recovers as the pressure increases, it exhibits a number of interesting and unique features. For example, the higher dispersion of Xe means that the gas pressure $P$ required for gain recovery, i.e. $\rho_{01} \sim 1$, can be much lower than in the case of pure hydrogen. For instance, $\rho_{01} > 0.9$ is achieved at a pressure of $\sim 38$ bar for pure hydrogen (upper panel in Fig. 1(c)) and $\sim 17$ bar for the mixture with $p_{H2} = 15$ bar (brown curve in the lower panel of Fig. 1(c)). Note that the slight reduction in gain at higher Xe pressures is due to a reduction in $T_2$ (Eq. (4)).

Another interesting feature is that $G_{01}^{TR}$ can actually be enhanced by the addition of a buffer gas. For instance, the maximum gain for $p_{H2} = 15$ bar, represented by the brown solid curve in Fig. 1(c), reaches a value comparable to that obtained with 25 bar of pure $H_2$ at $p_{Xe} \sim 3$ bar – a somewhat counter-intuitive result since the molecular density of $H_2$ does not change as $p_{Xe}$ increases. This is because the effective Raman gain (Eq. (4)) must be considered, not the material gain [8]. The addition of buffer gas frustrates coherent gain suppression by increasing the dephasing rate $\vartheta_{01}$, making it possible to recover a value close to that of the material gain ($\rho_{01} \sim 1$). This increase in the gain is even more pronounced at high pump intensities (see Fig. 1(b)) – an important feature in the design of fiber-based Raman shifters and Raman comb generators. Figure 1(d) also shows horizontal slices of Fig. 1(b) for two different pump intensities of 5, and 40 W/μm² in the upper and lower panel, respectively. These results show that addition of a noble gas allows control of the effective in-fiber Raman gain and thus the overall performance of the system.

In Fig. 2 the dephasing rate is plotted versus the buffer gas pressure $p_B$ for the same system as modeled in Fig. 1, when Xe, Ar or Ne are added, keeping $p_{H2}$ constant at 10 bar. The differing buffer gas dispersions [18] permit adjustment of the gain suppression pressure over a wide range – a feature that could be of great importance, for example, in the generation of single-mode anti-Stokes sidebands. Moreover, unwanted effects such as mutual collisional broadening and electronic resonances, especially in the ultraviolet region, can also be mitigated by choice of a suitable noble gas [7].

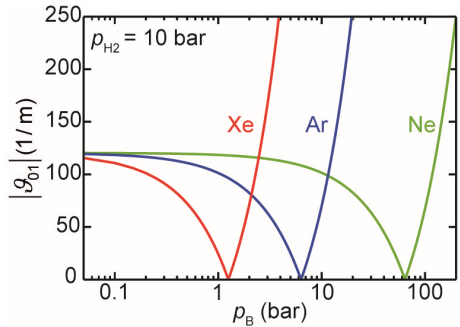

FIG. 2. Dephasing rate for different H₂/noble gas mixtures at $p_{H2} = 10$ bar and increasing $p_B$. The other parameters are that same as in Fig. 1.

To verify these predictions, we built the set-up in Fig. 3. A 37-cm length of the kagomé-PCF mentioned above was pumped with few μJ, 1 ns, pulses at 532 nm, generated by a microchip laser. The extremely thin core walls enabled operation in the ultraviolet since the first anti-crossing between the LP$_{01}$-like core mode and resonances in the core-wall lies at 210 nm [19].

To allow precise control of the partial pressures of the mixing gases, we filled each gas species into a separate gas bottle at the required pressure. These gases were then mixed in a separate bottle before being pumped into the fiber. This procedure was essential to ensure homogeneity of the gas mixture – if injected into the fiber separately, the large difference in diffusion rate between individual gases would cause non-uniform gas concentrations. In addition, we monitored the power of the pump, Stokes and anti-Stokes bands over time so as to ensure that the system had reached equilibrium before doing any final measurement. Two different sets of experiments were conducted, all of them at a pump pulse energy of 3.6 μJ. First, the fiber was filled with pure H₂ and the pressure varied from 5 to 35 bar (Fig. 4(a)). Second, the partial pressure of H₂ was fixed at 12 bar (Fig. 4(b)) and 13.5 bar (Fig. 4(c)) and $p_{Xe}$ was swept. The energy in each sideband and the near-field mode profiles were recorded while scanning the pressure in each case. The measured energies of the first Stokes (at 683 nm) and anti-Stokes (at 435 nm) bands are represented by the data points in Fig. 4. The results of numerical simulations (solid curves in Fig. 4) using a multimode-extended set of Maxwell-Bloch equations [15] show very good agreement. As reported in [8], the Stokes band tends to be emitted in an LP$_{11}$-like mode when parametric gain suppression is strong for the LP$_{01}$-like mode.

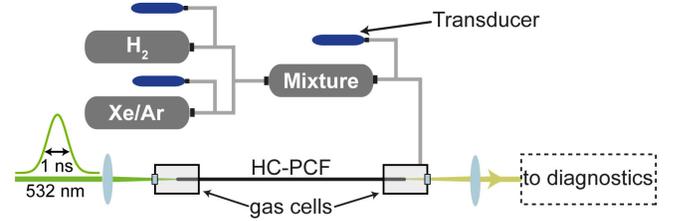

FIG. 3. Schematic diagram of the experiment. The pressure scan was performed in steps of ~100 mbar of buffer gas. Near-field images and sideband energies were recorded at each pressure point.

These results confirm that the behavior of the Stokes and anti-Stokes signals qualitatively follows that of the pure gas (Fig. 4(a)), but at a lower total pressure. In particular, the second anti-Stokes peak, which occurs at 22 bar for pure H₂, appears at 13.6 bar for a H₂/Xe mixture. It is also interesting that using a gas mixture reduces the strength of Stokes emission in the LP$_{11}$-like mode, at the pressure where conversion to the anti-Stokes band is maximum (see the near-field images in Fig. 4). The presence of the buffer gas causes a slight reduction in the material gain, while inhibiting the gain suppression, as explained above.

As a result, both Stokes and anti-Stokes bands are emitted in a pure LP$_{01}$-like mode at the point where the anti-Stokes signal is strongest, a situation that does not occur when there is no buffer gas. Remarkably, the conversion efficiency to the LP$_{01}$ anti-Stokes band, expressed as the ratio between the output energy in the sideband to the input pump energy, reaches 5%, a value which is, to our best knowledge, the highest yet reported in a single-mode fiber-based vibrational H₂ Raman convertor [10]. At $p_{H2} = 13.5$ bar, the Raman gain is even higher, resulting in ~40% conversion to the Stokes band, comparable to the values obtained with the pure gas at much higher pressures (~ 25 bar, although for pure H₂ the maximum Stokes emission occurs in a mixture of fiber modes), limited only by the short fiber length (Fig. 4(c)).

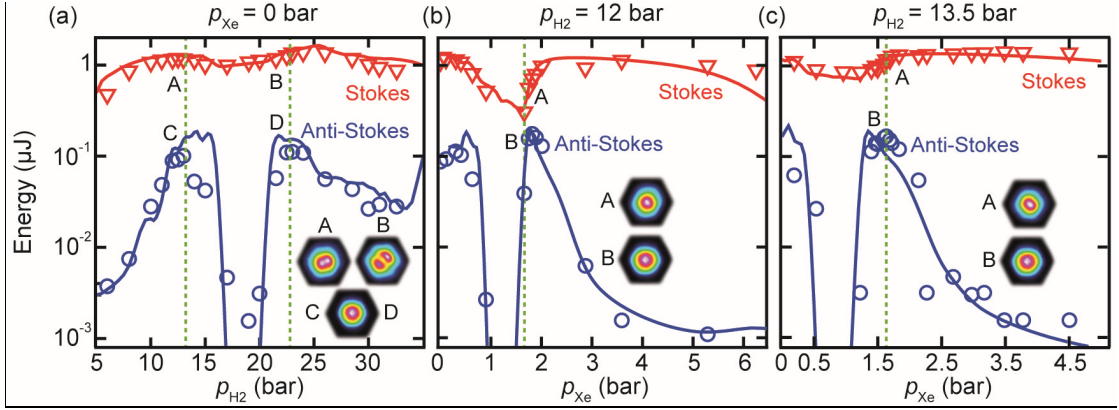

FIG. 4. Experimental (symbols) and simulated (solid lines) energy in the first Stokes and anti-Stokes bands for a pump energy of 3.6 μJ, plotted (a) against $p_{H2}$ for pure $H_2$, (b) against $p_{Xe}$ for $p_{H2}$ = 12 bar and (c) against $p_{Xe}$ for $p_{H2}$ = 13.5 bar. Also shown are near-field optical micrographs of the Stokes and anti-Stokes signals at pressures where the anti-Stokes emission is strongest (shown by the dashed-green lines).

The excellent agreement between simulation and experiment makes it possible to study numerically inaccessible aspects of the system. As an example, Fig. 5 shows evolution of the three signals along the fiber for 3.6 μJ pump energy. In Fig. 5(a) $p_{H2}$ = 22 bar, and in Fig. 5(b) $p_{H2}$ = 12 bar and $p_{Xe}$ = 1.6 bar. At these pressures, the dephasing rate is nonzero and the second peak of the anti-Stokes occurs. Both Stokes and anti-Stokes signals grow with a similar, moderate value of gain until the pump becomes substantially depleted and amplification ceases [20, 21]. In the vicinity of the gain-suppression pressure, however, a strong Stokes signal appears in the $LP_{11}$-like mode (red-dashed curve in Fig. 5(a)), causing pump depletion and drastically impairing $LP_{01}$ amplification. As stated above, the gain for intermodal SRS is reduced in the gas mixture, favoring Stokes generation in the $LP_{01}$ mode (see Fig. 5(b)).

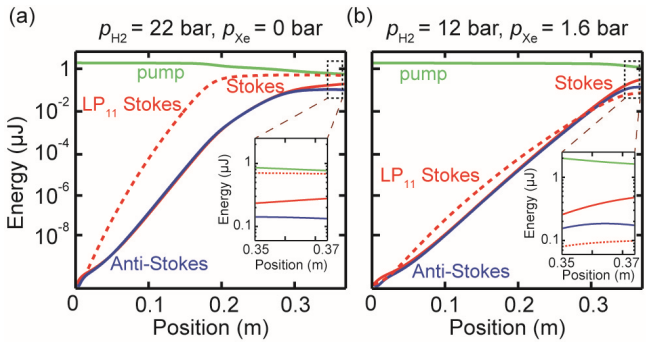

FIG. 5. Simulated evolution of the sideband energies with propagation distance at the maximum of anti-Stokes emission and 3.6 μJ pump energy for (a) $p_{H2}$ = 22 bar and $p_{Xe}$ = 0 and (b) $p_{H2}$ = 12 bar and $p_{Xe}$ = 1.6 bar. The parameters are otherwise identical to those in Fig. 4. The dashed red curve shows the $LP_{11}$-like Stokes signal, and the insets plot the evolution over the last 2 cm of fiber in both cases.

At higher pump energies (5.4 μJ in Fig. 6) the influence of the buffer gas is even more pronounced, with the maximum Stokes conversion clearly superseding that obtained with the pure gas and the pressure range, over which the $LP_{11}$ contribution to the Stokes band exceeds 10% (the shaded regions in Fig. 6(b)), shrinking significantly. The red curves are for $p_{Xe}$ = 0, and the blue curves for $p_{H2}$ = 12 bar and $p_{Xe}$ being varied. The improvement in Stokes conversion efficiency and $LP_{01}$ mode purity is evident. We also observe that the second-order Stokes band (at 953 nm) is emitted in the $LP_{01}$ mode, with a strength comparable to that obtained in the pure gas though at a total pressure a factor of two smaller. Interestingly, for $p_{Xe}$ = 0 the second Stokes signal is initially emitted in the $LP_{11}$ mode, gradually evolving to the $LP_{01}$ mode, just like the first Stokes signal. The same behavior is observed for the second anti-Stokes (at 368 nm), where the $LP_{01}$ conversion efficiency reaches ~2% [15], further demonstrating the excellent performance of the buffered-hydrogen-filled fiber for frequency conversion to the ultraviolet [22]. These results supersede those reported previously [4,5], where a buffer gas was used merely to dephase conversion to higher-order sidebands so as to concentrate conversion to the first Stokes.

A further interesting aspect of the PCF-based system is the onset of a strong backward Stokes signal for pure hydrogen at pressures >~20 bar when the pulse length is comparable with the fiber length (green curve in Fig. 6(a)) [23]. This effect is even stronger close to the gain suppression point, because the backward gain is not suppressed. Once again, by careful addition of buffer gas, the effective gain can be increased without increasing the material gain, resulting in a significant increase in threshold for backward SRS. Finally, in order to confirm the universality of this approach, we also performed experiments using mixtures of hydrogen and argon, finding excellent agreement with our predictions [15].

In conclusion, the combination of waveguide dispersion with gas mixtures offers a novel means of controlling the nonlinear optical response. In particular, judicious addition of a buffer gas can dramatically reduce the hydrogen pressure required for a given dynamics, increase the conversion efficiency and significantly enhance the effective Raman gain, as well as suppressing Stokes emission into the $LP_{11}$ mode and in the backward direction. Buffering can also enhance the generation of narrowband deep and vacuum ultraviolet signals [22], resulting in very high conversion efficiencies. Finally, preliminary studies suggest that single-pass anti-Stokes conversion efficiencies as high as 10% can be achieved if a pressure gradient is introduced along the fiber.

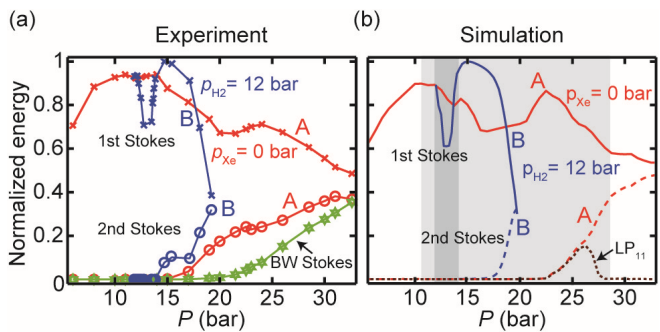

FIG. 6. (a) Measured energies in the first (crosses) and second (circles) Stokes signals, normalized to the overall maximum energy of the first Stokes for 5.4 μJ pump pulse energy. The red data-points (marked A) are for pure hydrogen and the blue (marked B) for a mixture with $p_{H2}$ fixed at 12 bar. The green data-points show the backward Stokes energy, measured for $p_{Xe} = 0$. (b) Results of numerical simulations, using the same labeling as in (a). Note that the simulations do not include backward SRS, which explains the disparity with experiment at high pressures of pure hydrogen. The shaded regions mark where Stokes emission into the $LP_{11}$ mode exceeds 10% of the total Stokes energy for pure gas (lighter shading) and a mixture (darker shading). The dashed brown line shows the second Stokes energy in the $LP_{11}$ mode in the case of pure gas.

# Supplementary Material

**Gain Reduction Factor**

The gain reduction factor $\rho_{ij}$, which reduces the strength of the coupling between the interacting fields, can be written [1]:

$$\rho_{ij} = \left| \text{Re}\sqrt{\left(\frac{q-1}{2}\right)^2 - \frac{\vartheta_{ij}}{g_P I_P}\left(\frac{\vartheta_{ij}}{g_P I_P} + i(q+1)\right)} \right| - \left(\frac{q-1}{2}\right), \quad \text{(S1)}$$

where $q = g_{AS}\omega_{AS}/g_P\omega_P$ and $\vartheta_{ij} = \beta_{AS}^{ij} + \beta_S^{ij} - 2\beta_P^{ij}$ is the dephasing rate, $\beta^{ij}$ being the propagation constant of the $LP_{ij}$-like core mode of a broadband-guiding HC-PCF [2]:

$$\beta^{ij} = \sqrt{k_0^2 n^2(p,\omega,T) - u_{ij}^2/a^2(\omega)}, \quad \text{(S2)}$$

where $k_0 = \omega/c$ is the vacuum wavevector, $c$ is the speed of light in vacuum, $u_{ij}$ is the $j$-th root of the $i$-th order Bessel function of the first kind, and $a(\omega)$ is the frequency-dependent effective core radius [2]. $n(p,\omega,T)$ is the refractive index of the filling gas at pressure $p$ and temperature $T$, which for a $M$-component gas mixture with partial pressures $p_k$ can be expressed as [3]:

$$n_{\text{mixture}}^2(\omega,p_1,...,p_M,T) - 1 = \sum_{k=1}^{M}\left(n_k^2(\omega,p_k,T) - 1\right) \quad \text{(S3)}$$

**Analytical solution of the Maxwell-Bloch equations**

The Maxwell-Bloch equations describing the evolution of the Stokes field $E_S$ in the stimulated regime can be written as [4]:

$$\left(\frac{\partial}{\partial z} + \frac{1}{c}\frac{\partial}{\partial t}\right)E_S(z,t) = -i\kappa_2 Q^*(z,t)E_P(z,t)$$

$$\frac{\partial}{\partial t}Q^*(z,t) = -\frac{1}{T_2}Q^*(z,t) + \frac{1}{4}i\kappa_1 E_P^*(z,t)E_S(z,t), \quad \text{(S4)}$$

where $E_P$ is the pump field, $Q$ is the Raman coherence and the coupling coefficients are defined as:

$$\kappa_1 = \sqrt{\frac{2c^2 g_P T_2 \varepsilon_0^2}{N\hbar\omega_S}}, \quad \kappa_2 = \frac{N\hbar\omega_S \kappa_1}{2\varepsilon_0 c}, \quad \text{(S5)}$$

where $g_P$ is the Raman gain of the bulk gas at the pump wavelength, $T_2$ the lifetime of the Raman coherence, $\varepsilon_0$ the vacuum permittivity, $N$ the molecular number density, $\hbar$ the reduced Planck's constant, and $\omega_S$ the Stokes frequency.

Note that all the quantities are expressed in SI units, whereas in [4] the parameters were in Gaussian units. Using the Laplace transformation method [5] and assuming a constant, square-wave pump pulse with duration $\tau_P$ (i.e., in the limit of no pump depletion) and a medium with length $z$ we can derive a full solution for the Stokes field:

$$E_S(z) = E_S(0) + \sqrt{\frac{g_P I_P z T_2^{-1}}{2}} E_S$$
$$\int_0^{\tau_P} e^{-T_2^{-1}(\tau_P - \tau')} \frac{I_1\left(\sqrt{2zg_P I_P T_2^{-1}(\tau_P - \tau')}\right)}{\sqrt{\tau_P - \tau'}} d\tau', \quad \text{(S6)}$$

where $I_1$ is the first-order modified Bessel function of the first kind. Closed form solutions of Eq. S6 exist in the high gain regime ($g_P I_P L \tau_P \gg T_2$) for two asymptotic cases [5]:

1) <u>Steady-state regime</u> ($\tau_P T_2^{-1} \to \infty$). Shifting the upper integration limit in Eq. S6 to infinity, we obtain:

$$E_S(z) \simeq E_S(0)e^{g_P I_P z/2} \quad \text{(S7)}$$

2) <u>Transient regime</u> ($\tau_P T_2^{-1} \ll 1$). We use the approximation of $e^{-A} \sim 1-A$ (for $A \ll 1$). Thus, we would have:

$$E_S(z) \simeq \frac{E_S(0)}{\sqrt{\pi}} \frac{e^{\sqrt{2g_P I_P z \tau_P T_2^{-1}} - \tau_P T_2^{-1}}}{\left(8g_P I_P z \tau_P T_2^{-1}\right)^{1/4}} \quad \text{(S8)}$$

Note that these calculations are consistent with the results reported in [4]. In practice the condition of transient operation can be relaxed to $g_P I_P L \gg \tau_P/T_2 = \hat{\tau}_P$ provided the duration of the Stokes pulse is shorter than $T_2$ [6]. We have verified that both conditions in Eq. 1 in the main text are satisfied for our experimental parameters. This is shown in Fig. S1, where we plot the overall Raman gain versus $\hat{\tau}_P$. For the pressure range used in the experiments (5 to 35 bar) and a pump pulse duration of 1 ns, $\hat{\tau}_P$ lies between 1 and 5, which is graphically represented by the region enclosed between the dashed lines in Fig. S1. For a fiber length of 30 cm and pump energies of >1 µJ, the values of the overall gain would be greater than 100 (indicated by the green shaded region in Fig. S1), justifying operation in the high-gain transient regime.

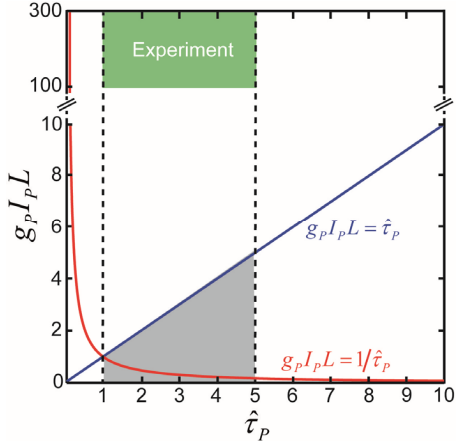

Fig. S1. The overall Raman gain plotted against $\hat{\tau}_P$ for a pump pulse with 1 ns duration. The blue and red curves show the limits for transient and high gain operation, respectively. In the experiments $(g_P I_P L)_{Exp} > 100$ and $\hat{\tau}_P$ lies in the region between the dashed lines (the green shaded region). This places us in the strongly transient regime.

To show the behavior of SRS in the different regimes, we plot in Fig. S2(a) the Stokes intensity using the above solutions in a typical case. The dashed-red curve is a plot of the full solution Eq. S6 versus $p_{H2}$ for a 1 ns-pump pulse centered at 532 nm – similar to that used in the experiments. The green and blue curves are plotted using the asymptotic expressions Eq. S7 and S8, respectively. The transient solution closely follows the general solution for the range of pressures investigated in the experiments, justifying the assumption that the system operates in the transient regime. At higher pressure the two curves differ more and more and the full solution approaches the steady-state ($T_2^{-1} \to \infty$), represented by the green-solid curve. With shorter pulse durations the conditions for transient behavior are more relaxed, with the result that the general solution follows the transient solution even beyond 60 bar.

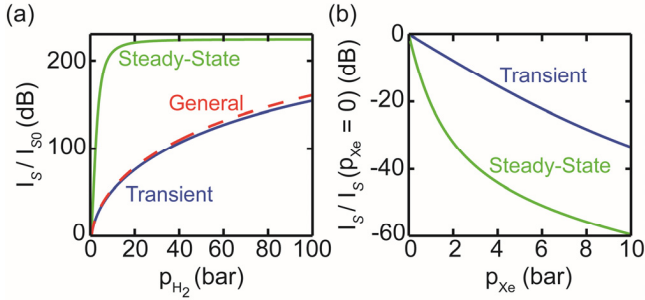

Fig. S2. (a) Normalized Stokes intensity versus the pressure of $H_2$ for parameters similar to those in the main text. The curves are plotted using the solutions in Eq. S6-S8 (for detailed discussion see text). (b) Stokes intensity plotted using Eq. S6 for a $H_2$/Xe mixture, keeping the partial pressure of $H_2$ constant. The blue curve shows the behavior of a typical system operating in the transient regime, and green curve a system operating in the steady-state regime.

In Fig. S2(b) we plot the Stokes intensity using the general solution in Eq. S6 for a case where the partial pressure of $H_2$ is kept constant and the partial pressure of the buffer gas $p_{Xe}$ varies from 0 to 10 bar. The blue curve is for a pump pulse duration of ~1 ns (transient-like behavior) and the green curve for a duration of ~10 ns (steady-state behavior). It is clearly seen that the reduction in Raman gain and Stokes intensity is less influenced by the presence of the buffer gas in the transient regime. Note that Fig. S2 is intended only for a qualitative comparison of the dynamics, which will be in practice more complex given the onset of other effects such as coherent gain suppression or amplification of higher-order sidebands. In particular, the effect of gain suppression can be included by replacing the material Raman gain $g_P$ with the effective Raman gain $\gamma_{ij}^{eff} = \rho_{ij} S_{ij} g_P$ in Eq. S6-S8 [1].

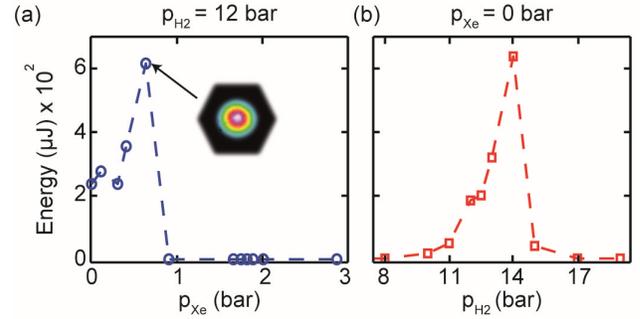

Fig. S3. Experimental (symbols) energy in the 2nd anti-Stokes sideband generated in (a) a $H_2$-Xe mixture, plotted versus $p_{Xe}$ when $p_{H2} = 12$ bar and (b) when $p_{Xe} = 0$ bar. A near-field image taken with a CCD camera is also shown in (a). The dashed connecting lines are just intended as guides for the eye.

## Numerical simulations

To simulate SRS dynamics in hollow-core photonic crystal fibers filled with mixtures of $H_2$ and Xe (Figs. 4, 5, and 6 in the main text), we employed a multimode-extended set of coupled Maxwell-Bloch equations [1]

$$\frac{\partial}{\partial z} E_{\sigma,l} = -\kappa_{2,l} \frac{\omega_l}{\omega_{l-1}} \sum_{\nu\xi\eta}^{M} i \frac{s_{\sigma\nu\xi\eta}}{s_{\nu\xi}} Q_{\nu\xi} E_{\eta,l-1} q_{\eta,l-1} q_{\sigma,l}^*$$
$$-\kappa_{2,l+1} \sum_{\nu\xi\eta}^{M} i \frac{s_{\sigma\nu\xi\eta}}{s_{\nu\xi}} Q_{\nu\xi}^* E_{\eta,l+1} q_{\eta,l+1} q_{\sigma,l}^* - \frac{1}{2} \alpha_{\sigma,l} E_{\sigma,l} \quad \text{(S9)}$$

$$\frac{\partial}{\partial \tau} Q_{\nu\xi} = -\frac{Q_{\nu\xi}}{T_2} - \frac{i}{4} s_{\nu\xi} \sum_l \kappa_{1,l} E_{\nu,l} E_{\xi,l-1}^* q_{\xi,l-1} q_{\nu,l-1}^*, \quad \text{(S10)}$$

where the integer $l$ denotes the sideband at frequency $\omega_l = \omega_P + l\,\Omega_R$ ($\Omega_R/2\pi = 125$ THz is the Raman frequency shift of the dominant vibrational mode of $H_2$) and the summations are over all possible permutations of the modal set $M$. We describe the complex electric field amplitude $e_{\sigma,l}(x,y,z,\tau) = F_\sigma(x,y) E_{\sigma,l}(z,\tau) q_{\sigma,l}$ of a guided mode (mode index $\sigma$) in terms of its normalized transverse spatial profile

$F_\sigma(x,y)$, the slowly varying complex field envelope $E_{\sigma,l}(z,\tau)$ and the phase term $q_{\sigma,l} = \exp[-i\beta_\sigma(\omega_l) z]$. $Q$ is the Raman coherence and $\tau$ is the time relative to a frame traveling at the group velocity of the pump pulse. The coupling constants $\kappa_{1,l}$ and $\kappa_{2,l}$ are defined in Eq. S5, and the overlap integrals $S_{\sigma\nu\xi\eta}$ and $S_{\nu\xi}$ in [1]. We also assume that the majority of Raman-active molecules remain in the ground state.

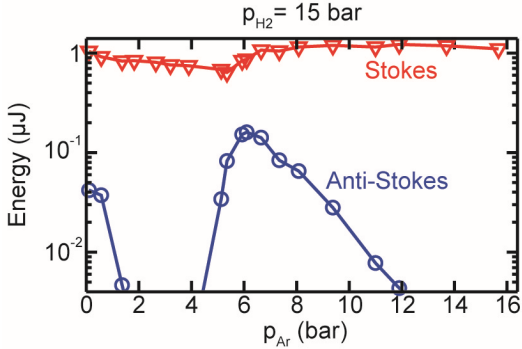

Fig. S4. Experimental (symbols) energy in the first Stokes and anti-Stokes for a pump energy of 3.60 μJ in a mixture of $H_2$ and Ar, plotted as a function of Ar pressure at $p_{H2}$ = 15 bar.

**Second anti-Stokes generation**

The second anti-Stokes energy (lying in the ultraviolet at 368 nm) generated in an experiment with a $H_2$/Xe mixture is shown in Fig. S3(a). The experimental parameters are the same as Fig. 4 in the main text (pump energy 3.6 μJ). The conversion efficiency is ~2%, which is similar to the case of pure $H_2$ shown in Fig. S3(b).

**Experiment with a mixture of $H_2$ and Ar**

As discussed in the main text, different noble buffer gases can be employed. To illustrate this we performed a similar set of experiments with argon, keeping $p_{H2}$ constant and varying $p_{Ar}$. This is shown in Fig. S4 for pump pulses with 3.6 μJ energy.

The level of the Stokes and anti-Stokes bands is also similar to that shown in Fig. 4(c) in the main text. Since argon is much less dispersive than xenon (see Fig. 2 in the main text), the pressure range over which the second anti-Stokes peak occurs is much broader than in the $H_2$/Xe mixture. This may be useful if the precise, relatively small pressure steps required for experiments with Xe are found technically challenging.